\documentclass[10pt,letterpaper]{article}
\usepackage[top=0.85in,left=2.75in,footskip=0.75in]{geometry}

\usepackage{amsmath,amssymb}

\usepackage{changepage}

\usepackage[utf8x]{inputenc}

\usepackage{textcomp,marvosym}

\usepackage{cite}

\usepackage{nameref,hyperref}

\usepackage[right]{lineno}

\usepackage{microtype}
\DisableLigatures[f]{encoding = *, family = * }

\usepackage[table]{xcolor}

\usepackage{array}

\newcolumntype{+}{!{\vrule width 2pt}}

\newlength\savedwidth

\raggedright
\usepackage[aboveskip=1pt,labelfont=bf,labelsep=period,justification=raggedright,singlelinecheck=off]{caption}

\makeatletter
\renewcommand{\@biblabel}[1]{\quad#1.}
\makeatother

\usepackage{lastpage,fancyhdr,graphicx}
\usepackage{epstopdf}
\fancyheadoffset[L]{2.25in}
\fancyfootoffset[L]{2.25in}
\usepackage[T1]{fontenc} 
\usepackage[utf8x]{inputenc}
\DeclareUnicodeCharacter{2212}{-}

\usepackage[ngerman]{datetime}
\newdateformat{myformat}{\THEDAY{. }\monthnamengerman[\THEMONTH] \THEYEAR}
\usepackage{parskip} 
\usepackage{svg}
\usepackage{import}
\usepackage{color}
\usepackage{float}
\usepackage{subfig}

\usepackage{tikz}
\usepackage{pgfplots}
\usepgfplotslibrary{external} 
 \tikzset{external/system call ={lualatex
 \tikzexternalcheckshellescape
 -halt-on-error
 -interaction=batchmode
 -jobname "\image" "\textsource"}}
\usepackage{amsmath} 
\usepackage{amssymb} %
\usepackage{wasysym} %
\usepackage{marvosym}
\usepackage{mathtools}
\usepackage{upgreek}
\usepackage{xcolor}
\usepackage{paralist}
\usepackage{enumitem} 
\usepackage{xcolor,hyperref}
\hypersetup{ 
	colorlinks=true,       
    linkcolor=blue,          
    citecolor=blue,        
    filecolor=magenta,      
    urlcolor=cyan           
    }
\usepackage[format=plain,
      justification=RaggedRight]
     {caption}
\newcommand{\subR}[1]{
  \protect\begin{NoHyper}\protect\subref{#1}\protect\end{NoHyper}
}

\newcommand{\mathsym}[1]{{}}

\newcommand{\e}{\mathrm{e}}

\newcommand{\LP}{\Delta}

\newcommand{\vc}[1]{\boldsymbol{#1}}

\newcommand{\partd}[1]{\partial_{ #1}}
\newcommand{\bra}[1]{\left( #1 \right)}
\newcommand{\brc}[1]{\left[ #1 \right]}
\newcommand{\bro}[1]{\left\{ #1 \right\}}

\newcommand{\g}[1]{\begin{gather}      #1     \end{gather}}

\newcommand{\alg}[1]{\begin{align}      #1     \end{align}}

\usepackage{comment}
\excludecomment{versionb}
\includecomment{versiona}

\begin{document}
\vspace*{0.2in}

\begin{flushleft}
{\Large
\textbf\newline{On biological flow networks: Antagonism between hydrodynamic and metabolic stimuli as driver of topological transitions.}
}
\newline

Kramer, Felix \textsuperscript{1,2},
Modes, Carl D. \textsuperscript{* 1,2,3},
\\
\bigskip
\textbf{1} Max Planck Institute for Molecular Cell Biology and Genetics (MPI-CBG),  Dresden 01307, Germany
\\
\textbf{2} Center for Systems Biology Dresden (CSBD),  Dresden 01307, Germany
\\
\textbf{3} Cluster of Excellence, Physics of Life, TU  Dresden, Dresden 01307, Germany
\\
\bigskip

%
%


\textcurrency Current Address: Center for Systems Biology Dresden (CSBD), Pfotenhauerstrasse 108 Dresden 01307, Germany 



* modes@mpi-cbg.de

\end{flushleft}
\section*{Abstract}

A plethora of computational models have been developed in recent decades to account for the morphogenesis of complex biological fluid networks, such as capillary beds.
Contemporary adaptation models are based on optimization schemes where networks react and adapt toward given flow patterns. Doing so, a system reduces dissipation and network volume, thereby altering its final form.
Yet, recent numeric studies on network morphogenesis, incorporating uptake of metabolites by the embedding tissue, have indicated the conventional approach to be insufficient. 
Here, we systematically study a hybrid-model which combines the network adaptation schemes intended to generate space-filling perfusion as well as optimal filtration of metabolites.
As a result, we find hydrodynamic stimuli (wall-shear stress) and filtration based stimuli (uptake of metabolites) to be antagonistic as hydrodynamically optimized systems have suboptimal uptake qualities and vice versa. 
We show that a switch between different optimization regimes is typically accompanied with a complex transition between topologically redundant meshes and spanning trees. 
Depending on the metabolite demand and uptake capabilities of the adaptating networks, we are further able to demonstrate the existence of nullity re-entrant behavior and the development of compromised phenotypes such as dangling non-perfused vessels and bottlenecks.

\section*{Author summary}
Biological flow networks, such as capillaries, do not grow fully developed and matured in their final and functional form.
Instead they grow a rudimentary network which self-organizes bit by bit in the context of their surrounding tissue, perfusion and other stimuli.
Interestingly, it has been repeatedly shown that this development is mechano-transductional in nature, coupling complex bio-chemical signaling to mechanical cues such as wall-shear stress.
In accordance with previous studies we propose a minimal hybrid model that takes into account stimuli in the form of the actual metabolite uptake of the surrounding tissue and the conventional wall-shear stress approach, and incorporate these into the metabolic cost function scheme. 
We present a numeric evaluation of our model, displaying the antagonistic interplay of uptake and shear stress driven morphogenesis as well as the topological ramifications and frustrated network formations, i.e. dangling branches, bottlenecks and re-entrant behavior in terms of redundancy transitions.

\linenumbers

\section{Introduction}
\label{sec:intro}
\textit{"Physiological organization, like gravitation, is a 'stubborn fact' and it is one task of theoretical physiology to find quantitative laws which describe organization in its various aspects."} - C.D. Murray, in \cite{Murray:1926tj}.
\\ 
\vspace{0.25cm}
The physiological organization referred to here by Murray nearly a hundred years ago is the morphogenesis of vascular networks. 
And true to his statement, Murray derived a theoretical framework based on simple hydrodynamic assumptions which would account for branching patterns emerging in the arteriole regime, now known as \textit{Murray's law}. 
By minimizing the overall shear stress for a volume constrained lumen of a model network he would determine the flow of scientific inquiry in this field for the rest of the century.
And yet, recent studies on metabolite transport in flow networks \cite{Meigel:2019fe,Meigel:2018hw,gavrilchenko2020distribution} have demonstrated, along with similar studies on flow homogeneity \cite{Chang:2019ip}, that purely shear stress based morphogenesis models of the capillary bed may be in dire need of revision.\\
Biological flow networks do not grow fully developed and matured in their final and functional form, but seem to self-organize bit by bit in the context of their surrounding tissue, perfusion and other stimuli over a long time span (compared to hydrodynamic time scales) \cite{DAVIES:1995bc,PMID:18802843}.
Most interestingly, it has been repeatedly shown that this development is mechano-transductional in nature, coupling complex bio-chemical signaling to mechanical cues such as wall-shear stress \cite{1997Natur.386..671R,auxin,Nguyen:2006kr,pries1995}. 
This has been demonstrated to be the case in a variety of  vertebrate model organisms \cite{Nguyen:2006kr,Lenard:2015de,DahlJensen:2018dba}, for endothelium and epithelium alike. 
Stress based adaptation seems universally present in the biosphere, as it has further been observed in plants \cite{RothNebelsick:2001dh} and slime mold \cite{Tero:2007hh}.
A recurring problem in these studies is to identify stimuli causing the complex topology of capillary networks, i.e. resilient, space-filling meshes containing hydrodynamically favorable 'highways', defying Murray's law\cite{Karschau:2020hf,caro_pedley_schroter_seed_parker_2011,Sherman:1981ux}.
Subsequently, more complex stimuli such as growth, noisy flow patterns and hemodynamical complications have been rigorously discussed in computational studies, and have been found to account for complex topological changes in the respective networks
\cite{2010Sci...327..439T,Grawer:2015bh,Ronellenfitsch:2016hh, Ronellenfitsch:2019fe,2012PLoSO...745444H,2017PLSCB..13E5892C}.\\
Though these abstract schemes have been very successful in accounting for the topological complexity of biological flow networks, most disregard the physiological aspect entirely: It is general consensus that any major exchange of metabolites between vasculature and tissue, such as oxygen, salts, glucose, proteins  etc., is performed on the capillary level and poses a valid stimuli for morphogenesis \cite{TN_libero_mab23719635,adair1990}.
 Subsequently, identifying remodeling mechanisms and efficient network architectures has become crucial  to address vasculature pathology, e.g. tumor angiogenesis \cite{Pries:2009fx,pmid33805699,Welter2016}, or the intelligent design of synthetic supply systems \cite{Grigoryan458,mi9100493}.
Yet, to our knowledge, only a handful of suggestions have been made for heuristic stimuli models ensuring perfusion of vascular networks with metabolites \cite{Pries:1998tn,Secomb:2013hm}.
These initial approaches have been lacking a thorough discussion of the actual uptake capabilities of the embedding tissue as well as an incorporation into the current discussion of flow network optimallity.
Recently two frameworks have been re-visited to address this inquiry: the Taylor dispersion model \cite{doi:10.1098/rspa.1953.0139} and the Krogh model \cite{Krogh:1919jfa}.
Taylor dispersion models consider metabolite transport to be a non-trivial interplay between the flow pattern and diffusion while incorporating a concentration dependent solute uptake along the vessel surfaces. 
This dispersion model was initiated as an appropriate metabolite transport model during slime mold morphogenesis \cite{2013PNAS..11013306A,Marbach:2016bf}, but has since been utilized to account for the adaptation phenomena found in plants and vertebrate capillaries \cite{Meigel:2019fe,Meigel:2018hw}.
On the other hand, Garvrilchenko et al \cite{gavrilchenko2020distribution} suggested an adaptation model on the grounds of the Krogh model to account for the explicit supply of discrete service elements in the tissue. 
This transport model also incorporates the convection of solute along idealized vessels in combination with metabolite diffusion from the channel surface into the surrounding tissue environment. \\
Inspired by with these studies we propose a minimal hybrid model that takes into account stimuli in the form of the actual metabolite uptake of the surrounding tissue together with the the conventional wall-shear stress approach, and incorporate these into the metabolic cost function scheme. 
We begin in section \ref{sec:theory} by giving a brief illustration of the model framework, including the Kirchhoff network approach and introduction of metabolic cost functions. 
Next we present the numeric results, displaying the antagonistic interplay of uptake and shear stress driven morphogenesis as well as the topological ramifications and frustrated network formations, such as dangling branches and bottlenecks, see section \ref{sec:results}. 
We conclude and discuss these results in section \ref{sec:discussion}, pointing out the limitations of our model and parameter space search as well as the emerging phenomena of redundancy re-entrant behavior and plexus dependencies.

\section*{Methods}
\label{sec:theory}
In this section we give a brief overview on the theoretical framework of fluid advection and metabolite transport Kirchhoff networks. For readers already familiar with these concepts we propose to skip forward to section \ref{sec:results}.
 
\subsection{Approximating capillary beds as Kirchhoff networks}\label{sec:basic_circuit}
We model capillary networks of interest as a composition of thin fluid conducting channels, abstracted as a graph of $E$ edges and $N$ vertices.
These vertices will represent braching points of the real biological system and are assumed to hold no fluid volume on their own. Each edge $e$ carries a current $f_e$ such that at any vertex the sum of all currents is equal to a nodal source $s_n$.
Further, every edge is characterized by a conductivity $K_e$ relating the flow to the potential difference $\LP p_e$ according to Ohm's Law.
One may write these relations using the graph's incidence matrix $B_{ne}$ as
\g{
	s_n=\sum_{e} B_{ne} f_e \label{eq:current_law_vec},\\
	f_e=K_e \LP p_e\label{eq:ohm_law_vec},
}
where \eqref {eq:current_law_vec} is also referred to as a 'Neumann boundary condition', fixing the peripheral currents.
The resulting equation system determines the networks flow and pressure landscape, for further details see \cite{1955PCPS...51..406P, TN_libero_mab214116274}.
Here, we consider a system of Hagen--Poiseuille flows, which allows to express the current as the volume flow rate $f=\frac{\pi R^4}{8\eta L}\LP p$ \cite{Landau:1959vq}.
This ansatz considers the approximation of  thin cylindrical vessels of radius $R$ and length $L$ being perfused laminarly by a fluid of viscosity $\eta$.
Such a vessel system can be directly described via the presented Kirchhoff network by setting the conductivity on each link as $K_e=\frac{\pi R_e^4}{8\eta L_e}$.
 We do not incorporate any Non-Newtonian fluid properties though they may arise, for example, in blood.
\subsection{Metabolite transport in lumped network models}\label{sec:metabolite uptake}
We next must incorporate the metabolite transport of a single species of molecules across the network.
To do so, we take an approach inspired by Heaton et al \cite{Heaton:2012dk}: 
Solute is transported by means of diffusion and advection along a a quasi-one dimensional vessel, neglecting lateral perturbations or any dependencies of the flow velocity on the concentration levels. 
While molecules are drifting down the channel they are absorbed at constant rate by the channel walls which form the periphery to the embedding tissue, see Figure \ref{fig:toy_model}. 
Subsequently we formulate the dynamics of the concentration $\bar{c}\bra{z,t}$ along the channel's symmetry axis (z-axis)  by using the continuity equation:

\begin{figure}[!ht]
\begin{versiona}
\includegraphics[scale=.35]{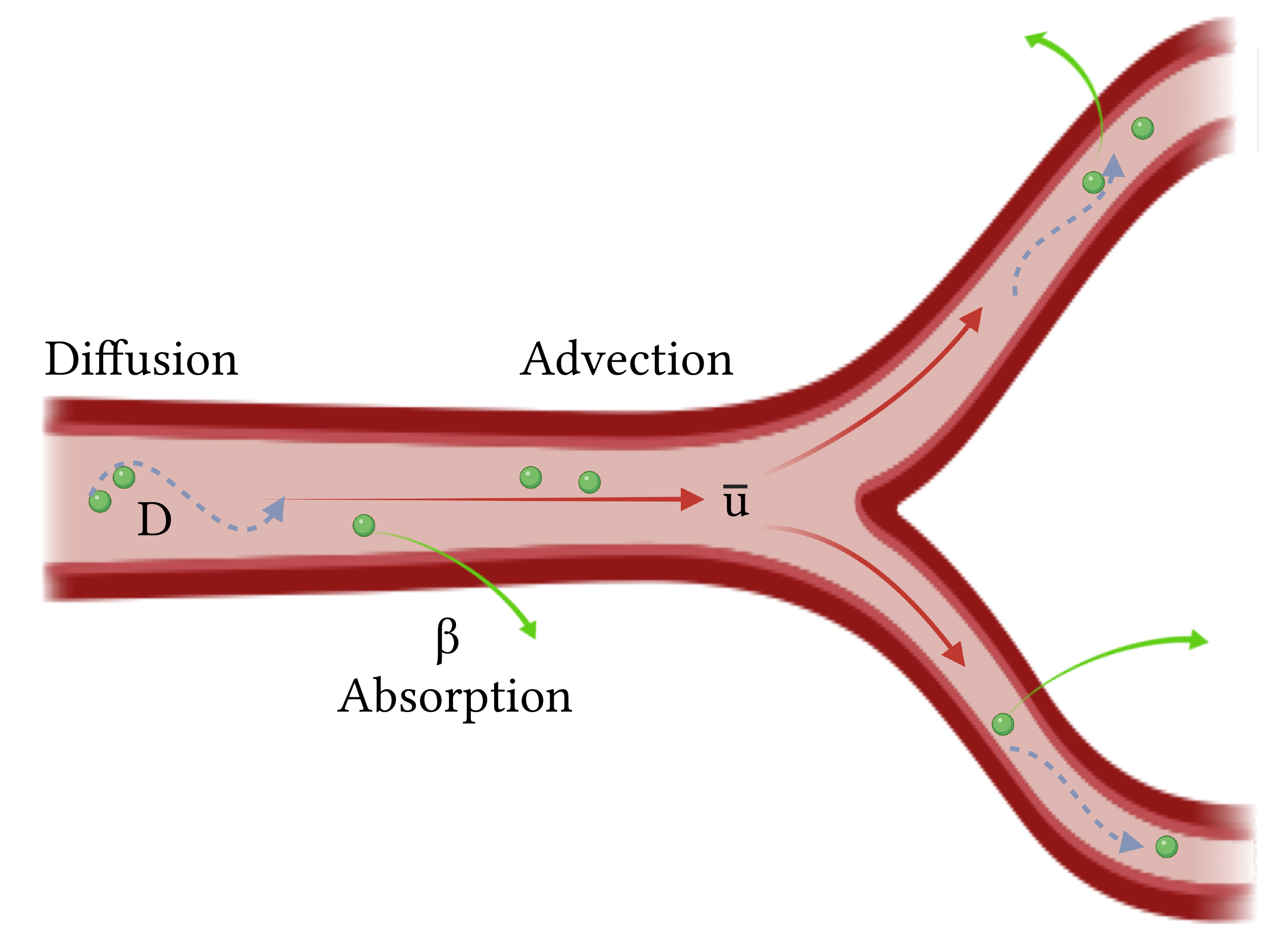}
\end{versiona}
\caption{
Toy model approach: Along the vessel system we assume a specific metabolite to diffuse along the concentration gradients, which are particularly impacted by the advection and surface absorption. \textit{(created with BioRender.com)} 
}\label{fig:toy_model}
\end{figure}

\alg{
	\partd{t}\bar{c}=D\partd{zz}\bar{c}-\bar{u}\partd{z}\bar{c}-\beta\bar{c}\label{eq:continuity}
}
Here, the diffusion constant is $D$, the advection velocity is $\bar{u}$ and the surface absorption rate is $\beta$.
 We use the notation $\bar{(\cdot)}$ to identify cross section averages, which denote the relevant factors in our deduction.
Though we will soon discuss the dynamics of an adapting network, we would like to point out here that the hydrodynamic time scales are considerably shorter than the time-scales of adaptation.
Therefore, we allow the system to reach a hydrodynamically stationary regime and calculate solutions only for $\partd{t}\bar{c}=0$. 
The resulting ordinary differential equation may readily be solved (using the notation $z^*=z/L$) such that we arrive at 
\alg{
	0= & \frac{D}{L^2}\partd{z^*z^*}\bar{c}-\frac{\bar{u}}{L}\partd{z^*}\bar{c}-\beta\bar{c} \label{eq:stat_concentration}\\
	\Rightarrow \bar{c}\bra{z^*}= & X_{0} \e^{ \lambda_{+} z^* }+X_{1}\e^{ \lambda_{-} z^*}\\
	\text{with }\lambda_{\pm}&=\frac{\bar{u}L}{2D}\pm\frac{\sqrt{\bra{\frac{\bar{u}L}{D}}^2+\frac{4\beta L^2}{D}}}{2}\\ &= \frac{1}{2}\bra{ Pe \pm\sqrt{Pe^2 + \beta^*  }}
}
One may deduce from the channels boundaries $\bar{c}\bra{z=0}=\bar{c}_0$, $\bar{c}\bra{z=L}=\bar{c}_L$ that $X_{0}=\frac{\bar{c}_L-\bar{c}_0\e^{a_{1}}}{\e^{a_0}-\e^{a_1}}$ and $ X_{1}=\frac{\bar{c}_0\e^{a_{0}}-\bar{c}_L}{\e^{a_0}-\e^{a_1}}$.
Here we define the Peclet number $Pe$ and the dimensionless absorption rate $\beta^*$ as: 
\alg{
	Pe &=\frac{\bar{u}L}{D}\\
	\beta^* &=\frac{4\beta L^2}{D}
}
Now we solve these equations for arbitrarily directed edges $e$, whose starting and end points we label as  $\alpha\bra{e}$, $\omega\bra{e}$. 
Hence we label all nodal concentrations as $\bar{c}_{e,0}=\bar{c}_{\alpha\bra{e}}$, $\bar{c}_{e,L}=\bar{c}_{\omega\bra{e}}$. 
Each link is further assigned its cross section area as $A_e=\pi R_e^2$, and we introduce the abbreviation $x_e=\sqrt{Pe_e^2 + \beta^*_e  }$.
Using this notation we may rewrite the concentrations $\bar{c}_k\bra{z}$ along each link $e$ as
\g{
\bar{c}_e\bra{z^*}=  \frac{\e^{\frac{Pe_e z^*}{2}}}{\sinh\bra{\frac{x_e}{2}}}\bro{\bar{c}_{\omega\bra{e}}\sinh\bra{\frac{x_e z^*}{2}} \e^{-\frac{Pe_e}{2}} \nonumber \right. \\ \left. - \bar{c}_{\alpha\bra{e}}\sinh\bra{\frac{x_e\brc{z^*-1}}{2}} }\label{eq:concentration_nodes}
}
Therefore we may calculate the solute flux on each link as 
\alg{
 I_e\bra{z} &= A_e \brc{ \bar{u}_e\bar{c}_e\bra{z} - D\partd{z}\bar{c}_e\bra{z} }\\&= q_e \brc{ Pe_e\bar{c}_e\bra{z^*} - \partd{z^*}\bar{c}_e\bra{z^*} }
 }
We use the notation $I_e\bra{0}=I_{\alpha\bra{e}},I_e\bra{L}=I_{\omega\bra{e}}$ and introduce the dimensionless flux parameter ${q_e=\frac{A_e D}{L}}$, see appendix S\ref{apx:metabolite_transport} for details.
Then we formulate the boundary conditions for the flux analogue to the Kirchhoff conditions of flow \eqref{eq:current_law_vec}. In particular we have balance of solute in- and outflux $J$ at each vertex as
\alg{
J_n & =\sum_{e\in out(n)} I_{\alpha\bra{e}} - \sum_{e\in in(n)} I_{\omega\bra{e}}
\label{eq:flux_boundary}
}
where $out(n)$, $in(n)$ indicate the index sets of arbitrarily directed edges pointing outwards or inwards of a vertex $n$.
We solve these equations according to a method described in \cite{1988PhRvA..37.2619K}:
Sorting the equations \eqref{eq:flux_boundary} for the concentration terms one can rewrite the entire equation system as
\g{ 
\vc{M}\cdot\vc{c}=\vc{J}\label{eq:fasym_system}
}
with an asymmetric matrix $\vc{M}$ whose entries are dependent only on $Pe$ and $\beta^*$, see appendix S\ref{apx:metabolite_transport} for details.
Considering a system with absorbing boundaries, one sets a subset of nodes $\vc{c}_0=\vc{0}$, while setting $J_n \geq 0$ for any node with $s_n \geq 0$, thereby matching the sources and solute influx vertices.  
Doing so one reduces the equation system \eqref{eq:fasym_system} and enables a unique solution for the remaining set of equations, determining the nodal concentrations $\bar{c}_n$ , see appendix S\ref{apx:metabolite_transport} for details. 
Utilizing these results one may calculate the total solute uptake per link $k$, as $\Phi_e=\beta L\int^1_0 \bar{c}_e\bra{z^*} dz^*$ and we get
\alg{
\Phi_e =q_e\bro{ \bar{c}_{\alpha\bra{e}}\brc{ x_e \coth\bra{\frac{x_e}{2}} - \frac{x_e\e^{\frac{Pe_e}{2}}}{\sinh\bra{\frac{x_e}{2}}} +Pe_e}  \right. \nonumber\\ \left. + \bar{c}_{\omega\bra{e}}\brc{ x_e \coth\bra{\frac{x_e}{2}} - \frac{x_e\e^{-\frac{Pe_e}{2}}}{\sinh\bra{\frac{x_e}{2}}} - Pe_e} }\label{eq:uptake_phi}
}
Note that the effective uptake $\Phi_e $ of a vessel is entirely determined by the landscape of Peclet numbers $Pe$ and local uptake rates $\beta$, see appendix S\ref{apx:metabolite_transport} for details. 
We shall capitalize on this phenomenon in the next sections to construct an adaption scheme which enables a regulation of effective metabolite uptake on the grounds of flow.
As a final remark, this approach would need to be extended in case of lateral perturbations in the concentration profile, i.e. $c\bra{z,r}=\bar{c}\bra{z} + \delta \bar{c}\bra{z,r}$. 
Here, the cross section average $\bar{c}\bra{z}$ would have to be considered in conjunction with a radial disturbance $\delta \bar{c}\bra{z,r}$, corresponding to the case of classic Taylor dispersion \cite{GITaylor}. 
Combining this framework with the absorption of metabolite was discussed in great detail by
Meigel et al \cite{Meigel:2018hw, Meigel:2019fe}.
As we here consider the small capillary regime, however, we assume that the effects of lateral concentration perturbations are negligible and refrain from considering them.

\subsection{Multi-target driven radius optimization}\label{sec:optimization}
In order to model the dynamic adaptation of rudimentary flow networks we follow the ansatz of characterizing such a transport system with a cost function, $\Gamma$. Any minimization of the cost $\Gamma$ may be formulated as the long-term radial and topological adaptations of the flow network, and therefore determine its pruning behavior. Hence one may formulate a generic cost for vessel systems as proposed before by Ref.\ \cite{Bohn:2007fi}:
\g{
	\Gamma=\sum_e \bra{ \frac{f_e^2}{K_e} +\alpha_0  K_e^{1/2} } \label{eq:cost_ansatz_generic},
}
where the first term $\bra{\frac{f_e^2}{K_e}}$ is the power dissipation of the flow and the second is a metabolic cost $\bra{\alpha_0  L_eK_e^{1/2}}$, with proportionality factor $a$. 
Minimizing the first term encapsulates the notion of reducing the overall wall shear stress imposed on the tubal cells \cite{Hu:2013io}. 
The second term formulates a constraint on the conductance the biological organism may deploy or sustain. In particular, for Hagen-Poiseuille flows this can be seen as a radial constraint, preventing arbitrarily large vessel volumes. 
These metabolic functions may be tailored for virtually any biological flow network \cite{Chang:2019ipa}. \\
In this study we consider a dissipation-volume minimizing system in combination with the metabolic needs of the surrounding tissue. 
To reduce the problem's complexity we will consider the special case of constant channel lengths throughout the system $L_e=L$.
Similar to previous setups studied in Refs. \cite{Krogh:1919jfa,Meigel:2018hw,Pries:1998tn} we will focus here on a vessel network embedded in a tissue environment, where each vessel is surrounded by the service volume it supplies. 
Generally speaking these service volumes are ensembles of cells forming a bulk environment, signaling affiliated vessels to adjust supply properly, e.g. by downstream signalling (via convection) or up-stream signalling (via conduction) \cite{PMID:18802843}.
Here, individual vessels are supposed to act as \textit{fair players} by adapting toward a specific metabolite need and not beyond that. This may correspond to cases were tissues are able to prevent over-saturation, e.g. actively avoid toxic overdosing. \\
We consider each vessel to be surrounded by tissue to which it supplies a metabolite, as described in \ref{sec:metabolite uptake}. 
Further, every such element of tissue  $i$  demands a basic influx of solute $\Phi_{0,i}$ possibly mismatching the current uptake $\Phi_e$ provided by the embedded vessels. 
In addition to the metabolic cost \eqref{eq:cost_ansatz_generic} we also define a mismatch cost $S\bra{ \vc{\Phi},\vc{\Phi}_0} \geq 0$ and write
\g{
	\Gamma=S\bra{ \vc{\Phi},\vc{\Phi}_0} + \sum_e \bra{ \alpha_1 \frac{f^2_e}{K_e} +\alpha_0 K_e^{1/2}}\label{eq:cost_ansatz_demand}
}
We construct $S\bra{ \vc{\Phi},\vc{\Phi}_0}$ in such a way that deviation from the demand $\vc{\Phi}_0$ is penalized. 
We minimize the metabolic cost \eqref{eq:cost_ansatz_demand} using a gradient descent approach where vessels are allowed to adjust their individual radii, which will alter the channel conductivities and subsequently the local Peclet numbers. 
The impact of the simultaneously given dissipation-volume constraints, as in \eqref{eq:cost_ansatz_generic}, is tuned via the coupling parameters $\alpha_0$, $\alpha_1$. 
Increased volume penalties $\alpha_0$ naturally lead to smaller vessel structures, simultaneously increasing $Pe$ and therefore hindering solute uptake.
On the other hand, increasing the dissipation factor $\alpha_1$ will generally increase vessel size for perfused vessels and decrease local $Pe$, thereby increasing uptake.

\subsection{Measuring topological network redundancy and metabolite filtration}
\label{sec:orderparameters}
In order to quantify the topological changes occurring in an adapting system, we monitor the amount of cycles in a network as \cite{Whitney:1932bg},
\g{
	Z= E - \bra{N-1} \label{eq:cycles}.
}
We use this metric in the following way: The network is first initialized for a densely reticulated system, a so called plexus. 
Links are no longer updated when their radius falls below a critical threshold $r_c$. We call such edges \textit{pruned} which corresponds to the biological phenomenon of having a vessel degenerate and collapse. 
At the end of each optimization we remove all pruned edges and disconnected vertices from the graph and recalculate the remaining number of cycles.
 We then define the relative nullity of an equilibrated network,
\g{
\varrho=\frac{E-N+1}{Z_0}
} 
as an order parameter, where $Z_0$ is the initial number of cycles before adaptation. 
Hence $\varrho=0$ corresponds to a treelike network while $\varrho>0$ captures the relative amount of redundancy in comparison to the initial plexus. 
Further we monitor the relative solute uptake of the network as
\g{
	\sigma = \log_{10}\brc{ \frac{\sum_e \Phi_e}{ \sum_{n, J_n \geq 0} J_n } }
}
Hence we get $\sigma \rightarrow 0$ if the network's vessel surface absorbs the entire injected solute and $\sigma \rightarrow -\infty$ if there is no absorption whatsoever. 
Generally we intend to construct the mismatch $S\bra{ \vc{\Phi},\vc{\Phi}_0} $ such that a certain value of $\sigma$ is reached, reflecting the notion that the supplied tissue needs to meet a certain total solute demand.


\section{Simulating demand-supply adaptation in capillary beds}\label{sec:results}
\begin{figure}[!ht]
 \centering
 \begin{versionb}
\def\svgscale{0.16}
\subfloat{
\label{fig:setup_1_links}
}\\
\def\svgscale{0.12}
\subfloat{
	\label{fig:setup_2_links}
	}
	\end{versionb}
	\begin{versiona}
	\def\svgscale{0.16}
\subfloat[]{
\hspace{-0.5cm}
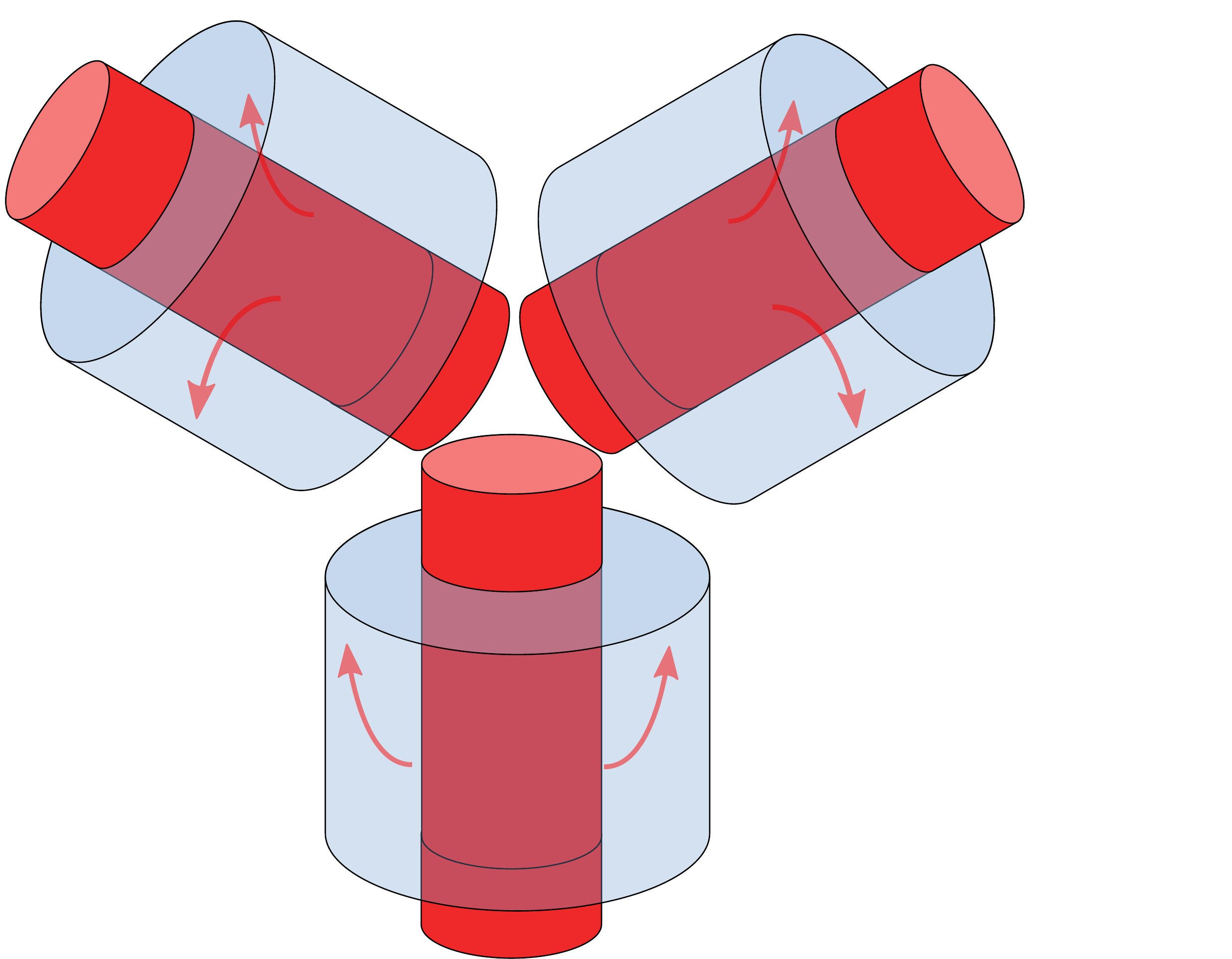\hspace{-0.75cm}
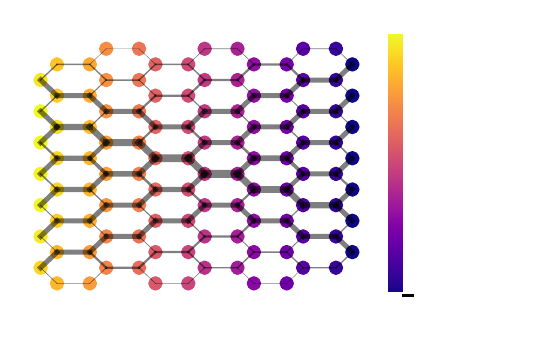
\label{fig:setup_1_links}
}\\
\def\svgscale{0.12}
\subfloat[]{
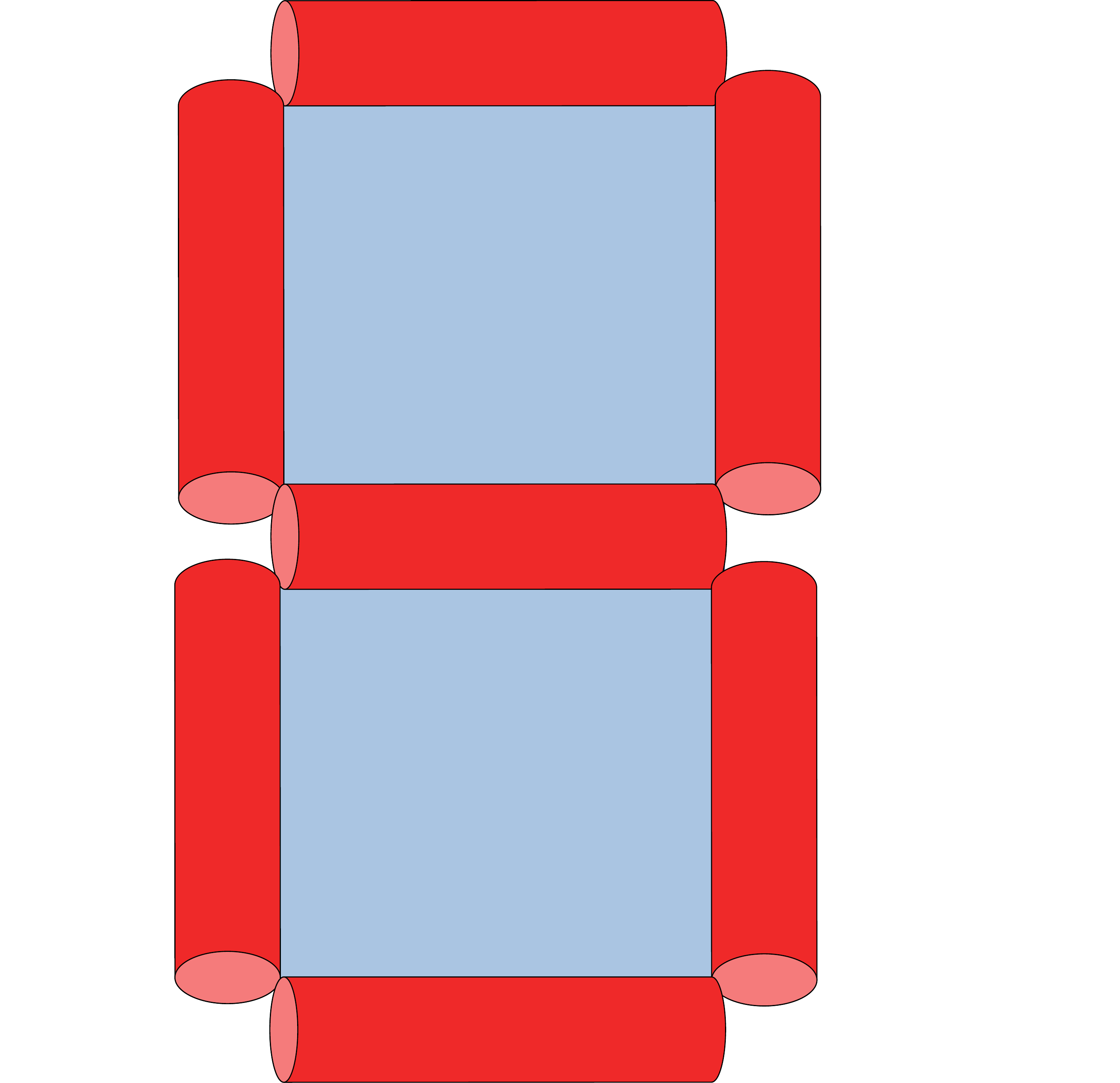\hspace{-0.5cm}
	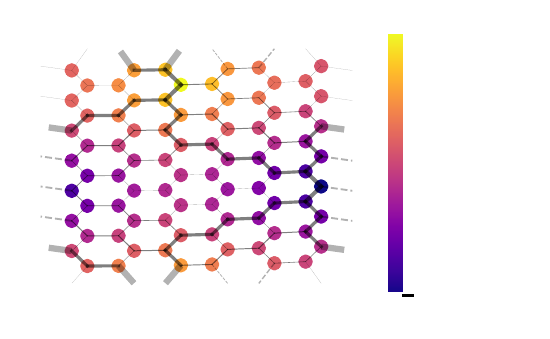	
	\label{fig:setup_2_links}
	}
	\end{versiona}
\caption{
Individual and collective uptake approaches and network plexi (with edge thickness representing radii and nodal concentrations color-coded):
\subR{fig:setup_1_links} Link-wise demand model, with edge-wise demand $\Phi_{0,e}$ and responding supplies $\Phi_{e}$. 
The planar plexus is initialized with multiple sources and absorbing boundaries on opposing graph sides. 
\subR{fig:setup_2_links} Volume-demand model, with enclosed tiles having a demand $\Phi_{0,v}$ responded to by affiliated edges with supply $\Phi_{e}$.
The plexus is initialized with periodic boundaries and a dipole source-sink configuration. 
Transparent, marked links indicate periodic boundaries.
}\label{fig:setup}
\end{figure}

In this section we present the simulation setup and numeric results as well as the implications of the hybrid model framework presented in the previous sections.
In particular we will interrogate Eq. \eqref{eq:cost_ansatz_demand} for two uptake scenarios: Link-wise demand-supply and volume-wise demand-supply, see Figure \ref{fig:setup}. For each we define individual mismatch functions $S\bra{ \vc{\Phi},\vc{\Phi}_0} $
\label{sec:solute_cost}.
We focus in this study on radial adaptation in order to adjust for unfavorable flow patterns. In particular we minimize the metabolic cost \eqref{eq:cost_ansatz_demand} using a gradient descent approach identifying local minima for random plexus initializations:
\g{
	\partd{t} r_i=-\chi\partd{r_i} \Gamma\\
   \Rightarrow \frac{d}{dt}\Gamma = \nabla \Gamma ^T \cdot \partd{t}\vc{r}  \leq 0 \text{, }\forall \chi \geq 0\nonumber
}
In appendix S\ref{apx:demand_supply_adaptation} we give a detailed account of the derivation of the respective dynamical systems for $\Gamma$, which defines local equations of motion for vessel radii. 

\subsection{Linkwise demand-supply model}
\label{sec:supply_demand_link}
\begin{figure}[!ht]
\begin{versiona}
\hspace{-0.8cm}
\includegraphics[scale=1]{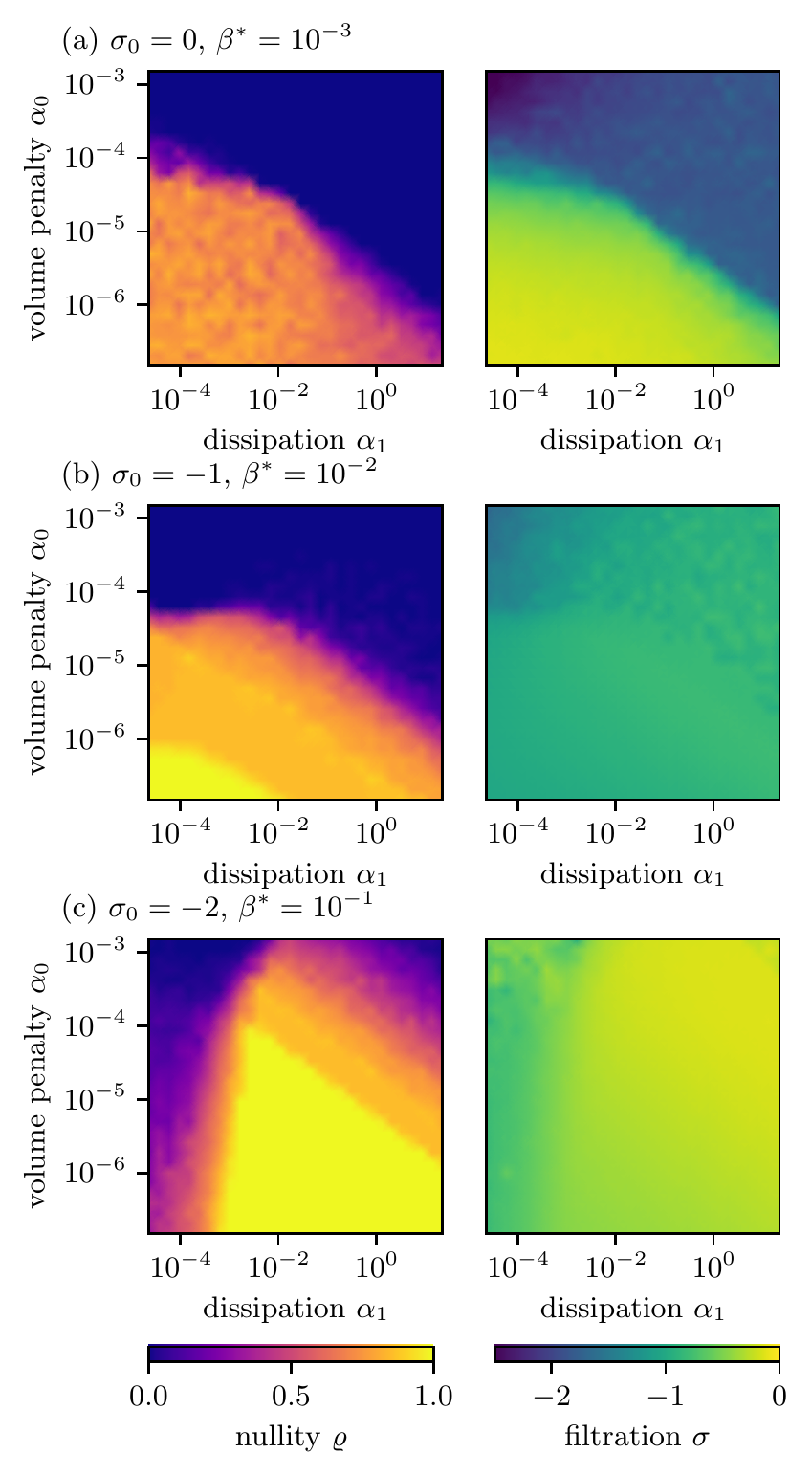} 
\end{versiona}
\caption{Stationary states for the link-wise demand-supply problem \eqref{eq:cost_system_1}: Color-coded state diagrams indicate the network nullity $\varrho$ (left column) and accumulated filtration $\sigma$ (right column) for selected $\bra{\sigma_0,\beta^*}$.
\label{fig:demand_supply_screen}
}
\end{figure}

\begin{figure}[!ht]
\begin{versionb}
\centering
\subfloat{ \label{fig:demand_supply_screen1}}\\
\subfloat{\label{fig:demand_supply_screen2}}\\
\subfloat{\label{fig:demand_supply_screen3}}
\end{versionb}
\begin{versiona}
\subfloat[$\sigma_0=0$, $\beta^*=10^{-3}$, $\alpha_0= 4.5 \cdot 10^{-5}$]{
\includegraphics[scale=1]{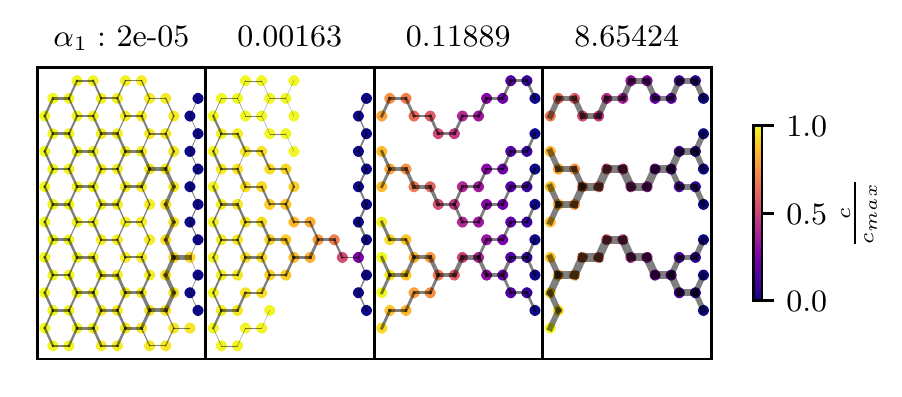} \label{fig:demand_supply_screen1}}\\
\subfloat[$\sigma_0=-1$, $\beta^*=10^{-2}$, $\alpha_0= 4.5 \cdot 10^{-5}$]{
\includegraphics[scale=1]{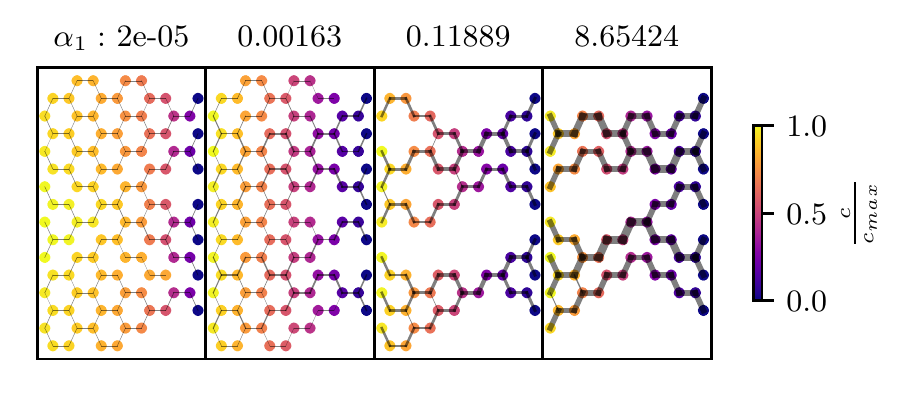} \label{fig:demand_supply_screen2}}\\
\subfloat[$\sigma_0=-2$, $\beta^*=10^{-1}$, $\alpha_0= 8\cdot 10^{-4}$]{
\includegraphics[scale=1]{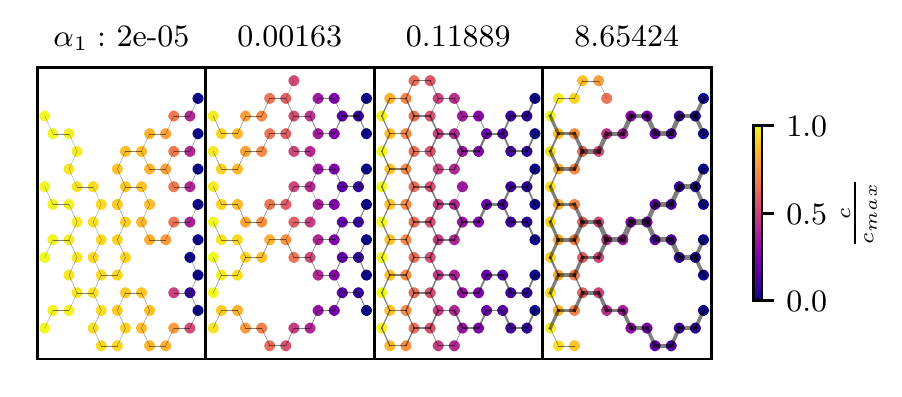} \label{fig:demand_supply_screen3}}
\end{versiona}
\caption{Stationary network formations for the linkwise demand-supply problem \eqref{eq:cost_system_1}: Depicted are formations for selected demands $\sigma_0$, absorption rates $\beta^*$, volume penalties $\alpha_0$ and dissipation feedbacks $\alpha_1$ after minimizing \eqref{eq:cost_system_2}. All systems were initialized with peripheral sink-source vertices. The nodal concentrations are color-coded and link-thickness is indicating edge radii.}\label{fig:demand_supply_screen_networks}
\end{figure}

We first consider a vessel system as shown in Figure \ref{fig:setup_1_links}, where each vessel supplies exactly one service volume.
Each such service volume demands an optimal influx of solute $\Phi_{0,e}$ possibly mismatching the current uptake $\Phi_e$ provided by the embedded vessel. 
The metabolite transport is computed as discussed in section \ref{sec:metabolite uptake}.
We formulate the uptake mismatch as a cost that reads:
\g{
S\bra{ \vc{\Phi},\vc{\Phi}_0}= \sum_e \bra{ \Phi_e - \Phi_{0,e}  }^2 
}
Hence we write for the system's total cost \eqref{eq:cost_ansatz_demand},
\g{
	\Gamma= \sum_e\brc{ \bra{ \Phi_e - \Phi_{0,e}  }^2 + \alpha_1 \frac{f^2_e}{K_e} +\alpha_0 K_e^{\frac{1}{2}}}\label{eq:cost_system_1}
}
In fact if the service volume demands are identical for the entire network we recreate the optimization framework studied in \cite{Meigel:2018hw} extended by wall-shear stress driven pruning. 
We impose sources $s_n >0$ and solute influx $J_n >0$ on all vertices of one side of the graph and sinks and absorbing boundaries on the opposing side. We refer to these vertices also as terminal or peripheral nodes. Internal vertices are set source- and influx-free, i.e. $s_n=0$, $J_n=0$.
In this study we are concerned with four essential model parameters: absorption rate $\beta^*$, demand $\phi_0$, dissipation feedback $\alpha_1$ and volume penalty $\alpha_0$.
We initialize the system for selected absorption rates $\beta^*$ and demand $\phi_0$ combinations, while scanning systematically for wide ranges of the dissipation feedback $\alpha_1$ and the volume penalty $\alpha_0$.
We set all vessels to correspond to a demand $\phi_{0,e}$ such that the network's demanded filtration rate corresponds to 
\g{
\sigma_0= log_{10}\brc{ \frac{\sum_e \phi_{0,e}}{\sum_{v, J_v>0}J_v}}.
} 
For the presented simulations, we initialize $\phi_{0,e}$ homogeneously across the network, as well as $\beta^*$. From here on we will discuss the demand in terms of the total network's demand $\sigma_0$. 
Following the adaptation algorithm, as described in the previous section, we find the system's stationary states and analyze those for their nullity $\varrho$ and actual filtration rate $\sigma$.\\
In diagram \ref{fig:demand_supply_screen} and \ref{fig:demand_supply_screen_networks} we display these metrics as well as stationary network formations for three significant $\bra{\sigma_{0},\beta^*}$ variations:
The first case, $\sigma_0=0$ and $\beta^*=10^{-3}$, depicted in Figure \ref{fig:demand_supply_screen}a and \ref{fig:demand_supply_screen1}, represents the unfavorable case of high demand paired with low absorption capability, prone to undersupply. 
As depicted in \ref{fig:demand_supply_screen1} we can show that increasing $\alpha_1$ will generally result in a nullity transition, displaying frustrations for the reticulated case as well as the formation of dangling branches not connected to any sinks.
Nevertheless we find the reticulated states in good agreement with a network wide adjustment toward the demand $\sigma_0$.
This seems counter-intuitive at first, as low  $\beta^*$ impair individual vessels from absorbing any significant amount of solute.
For small $\alpha_1$ we observe that the majority of vessels in the network are dilated while most peripheral connections to the sinks are degenerated and seemingly near to collapse, see appendix S\ref{apx:extra_plots}. 
Hence by dilating the  bulk of vessels one minimizes the Peclet numbers $Pe$ in the system ( boundary conditions dictate a constant volume throughput), which maximizes individual vessel uptake.
In order to guarantee high filtration, as many vessels as possible have to stay open, naturally resulting in a reticulated network state.
Note, that no significant concentration gradient is present in the bulk, see appendix S\ref{apx:extra_plots}.
At the periphery, incident to outlet nodes, decreasing the size of vessels leads to a sudden increase of the Peclet number $Pe$ and allows for rapid solute clearance in accordance with the boundary conditions.
These small vessels experience dramatically higher wall-shear stress than the rest of the system, but due to small $\alpha_1$ no significant growth feedback is posed. 
Subsequently, increasing $\alpha_1$ breaks this patterning, as it will open up exactly these high-shear stress vessels, which subsequently decreases $Pe$ at the periphery. 
This in turn changes the concentration landscape into a network wide gradient.
We further observe the break down of weakly perfused vessels, which are not anymore stabilized by the uptake mechanism.
See that this hybrid model stabilizes non-perfused branches (solute transport is administered via diffusion only) against volume penalty given an increase to the solute uptake feedback.

The second case, $\sigma_0=-1$ and $\beta^*=10^{-2}$ depicted in Figure \ref{fig:demand_supply_screen}b and \ref{fig:demand_supply_screen2}, features reduced demand paired with increased absorption capability. 
Here we see once again a nullity breakdown due to increase of $\alpha_0$ and $\alpha_1$, yet the corresponding filtration diagrams and concentration profiles are considerably different.
As shown in Figure \ref{fig:demand_supply_screen2}, even for marginal $\alpha_1$ a system spanning concentration gradient is readily abundant, see also appendix S\ref{apx:extra_plots}, and no degeneration of peripheral vessels takes place, as previously observed.
The filtration diagrams \ref{fig:demand_supply_screen2} illustrate that a seeming match of metabolite uptake is generally achieved in good approximation, yet deviates for large volume penalties $\alpha_0$.
Further, we see that dissipation feedback dominated regimes, where spanning trees emerge as the distinct graph topology, still reasonably well fulfill the initial filtration demand.
It seems that the current demand $\sigma_0$ can be met for the chosen absorption level $\beta^*$ with various network topologies indicating that the formation of hydrodynamically suitable networks might in general be possible.
From a filtration point of view, reticulation is not necessary, even though one should keep in mind that a match in overall filtration $\sigma$ does not necessarily imply an actual match of all absorbing vessels, nor does it ensure laterally extended supply.\\
The third case, $\sigma_0=-2$ and $\beta^*=10^{-1}$ depicted in Figure \ref{fig:demand_supply_screen}c and \ref{fig:demand_supply_screen3}, poses another unfavorable scenario where low demand is paired with high absorption rates, prone to oversupply.
Naturally we should observe general vessel degeneration in order to increase $Pe$, which in turn diminishes solute uptake.
Subsequently we would operate in a system that minimizes the number of vessels and experiences high wall-shear stress for the remaining vessels in the network.
That is indeed the case and may be observed for the network plots in Figure \ref{fig:demand_supply_screen3} at low $\alpha_1$. 
Nevertheless we observe the emergence of re-entrant behavior in the nullity diagram in case of increasing dissipation feedback $\alpha_1$.
Though the sharpness of this transition is sensitive to the choice of $r_c$ we find this behavior to be reliably reproducible for Neumann boundary conditions and to our knowledge has not been previously been encountered in other network morphogenesis models. 
As the phenomenon of re-entrant phase behavior can be caused by underlying antagonistic interactions, we hypothesize the transition to occur in the following way:
Approaching the nullity transition from the left-hand flank for small $\alpha_1$, we find the system 'loaded' with high wall-shear stresses as high Peclet numbers are abundant with a minimal set of channels which are not allowed to grow and redistribute load due to the small dissipation feedback.
Increasing $\alpha_1$ further, one is able to stabilize previously collapsing vessels and the system encounters a nullity transition. This process is still overshadowed by the fact that Peclet numbers have to be kept high by reducing overall vessel sizes.
Increasing $\alpha_1$ pushes the system once again toward the wall-shear stress dominated regime, where the adaptation stimulus for solute uptake becomes negligible.
Subsequently we observe the emergence of large conducting channels and a topological transition back towards spanning trees.
Though we find generally poor adjustment of the system toward low filtration rates we observe this relaxation to push the filtration rate up, as Peclet numbers decrease, see Figure \ref{fig:demand_supply_screen}.

\subsection{Volume-element demand-supply model}
\label{sec:supply_demand_volume}
\begin{figure}[!ht]

\begin{versiona}
\hspace{-0.7cm}
\includegraphics[scale=1]{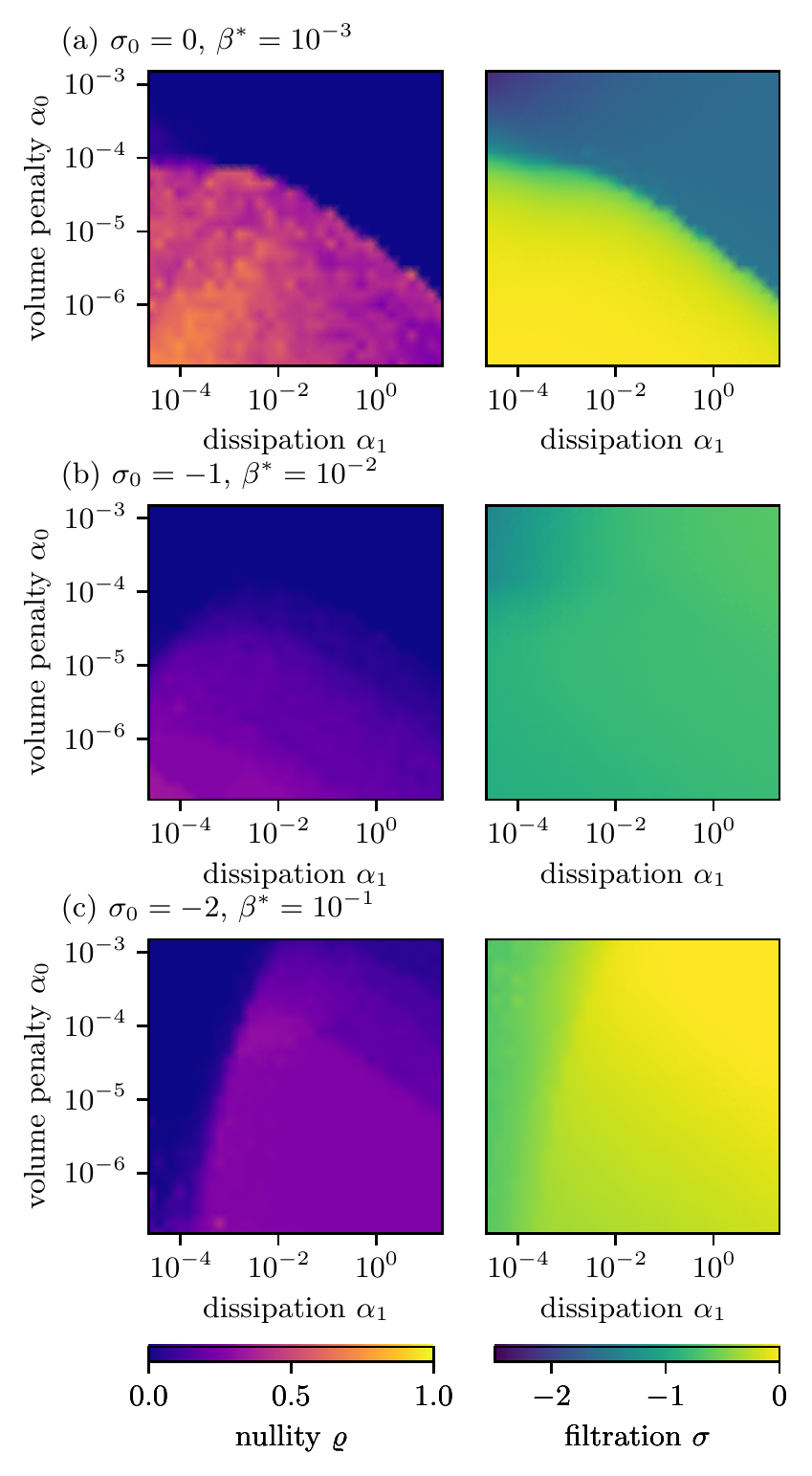}
\end{versiona}

\caption{Stationary states for the volume-wise demand-supply problem \eqref{eq:cost_system_2}: Color-coded state diagrams indicate network nullity $\varrho$ (left column) and accumulated filtration $\sigma$ (right column) for selected $\bra{\sigma_0,\beta^*}$.
}
\label{fig:volume_supply_screen}
\end{figure}

\begin{figure}[!ht]
\centering
\begin{versionb}
\subfloat{\label{fig:volume_supply_screen1}}\\
\subfloat{ \label{fig:volume_supply_screen2}}\\
\subfloat{ \label{fig:volume_supply_screen3}}
\end{versionb}
\begin{versiona}
\subfloat[$\sigma_0=0$, $\beta^*=10^{-3}$, $\alpha_0=2.4 \cdot 10^{-5}$]{
\includegraphics[scale=1]{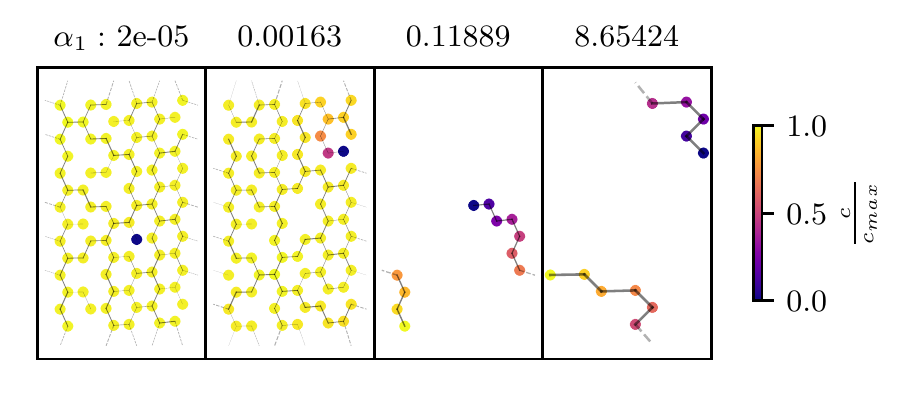} \label{fig:volume_supply_screen1}}\\
\subfloat[$\sigma_0=-1$, $\beta^*=10^{-2}$, $\alpha_0=2.4 \cdot 10^{-5}$]{
\includegraphics[scale=1]{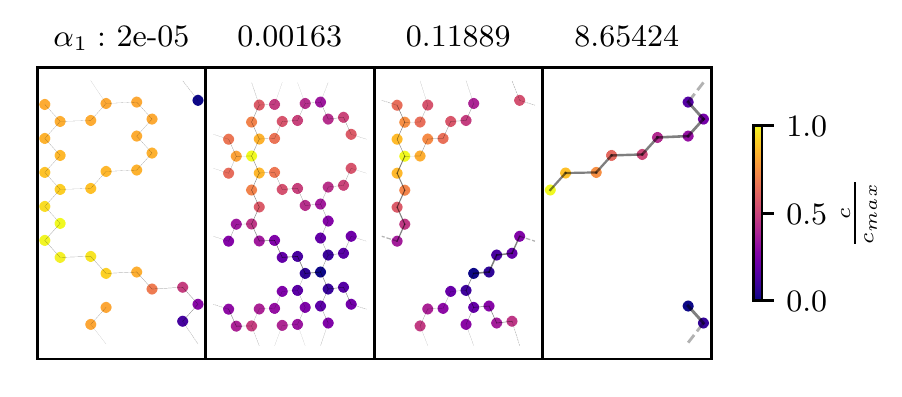} \label{fig:volume_supply_screen2}}\\
\subfloat[$\sigma_0=-2$, $\beta^*=10^{-1}$, $\alpha_0= 10^{-3}$]{
\includegraphics[scale=1]{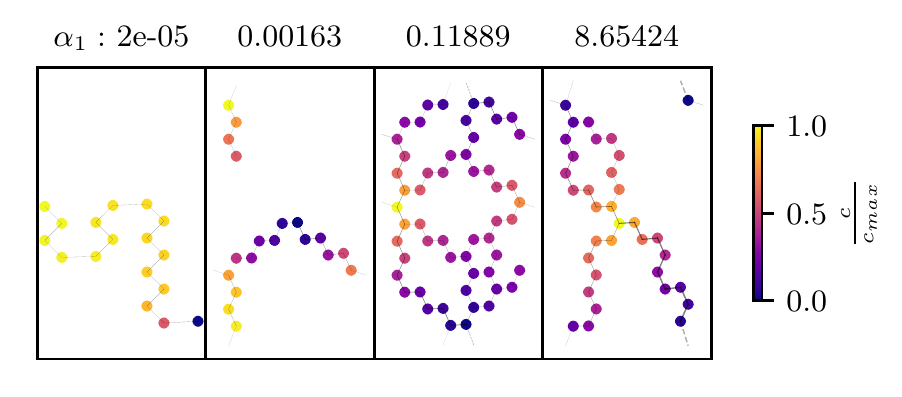} \label{fig:volume_supply_screen3}}
\end{versiona}
\caption{Stationary network formations for the volume-wise demand-supply problem \eqref{eq:cost_system_2}: Depicted are formations for selected demands $\sigma_0$, absorption rates $\beta^*$, volume penalties $\alpha_0$ and dissipation feedbacks $\alpha_1$ after minimizing \eqref{eq:cost_system_2}. All systems were initialized with a single sink-source dipole and periodic boundaries. The nodal concentrations are color-coded and link-thickness is indicating edge radii.}\label{fig:volume_supply_screen_networks}
\end{figure}

So far we have shown that demand-supply models are capable of rich phase behavior due to the antagonistic interaction of hydrodynamic and metabolic adaptation. 
Yet the current approach dictates only one supplying vessel for any service element, meaning that pruned vessels result in entirely non-supplied regions. Thus, in a real system no plexus could ever be pruned without killing multiple service elements.
In this section, we alter the previous setup and focus our studies on a system defined by shared volume elements rather than single vessel service volumes. 
Here, each service volume element $v$ is in contact with a set of vessels supplying individually a fraction of their uptake $\Phi_e$, see Figure \ref{fig:setup_2_links}. 
Each volume $v$ demands an influx of solute $\Phi_{0,v}$ potentially mismatching the summed uptake $\sum_{e \in E_v}\Phi_e$, provided by the attached vessel set $E_v$.
We do so following a similar model by Gavrilchenko et al \cite{gavrilchenko2020distribution}, which allows us to introduce redundancy and cooperative metabolite supply in the overall demand-supply scheme.
Similar to the previous section's setup, we propose a cost for volume-service of the form:
 \g{
 S\bra{ \vc{\Phi},\vc{\Phi}_0}= \sum_v\bro{  \Phi_{0,v} - \sum_{e \in V } \Phi_e   }^2
 }
 Hence we formulate the system's overall metabolic cost function:
\g{
	\Gamma=\sum_v\bro{  \Phi_{0,v} - \sum_{e \in E_v } w_e \Phi_e   }^2 + \sum_e \bra{ \alpha_1 \frac{f^2_e}{K_e} +\alpha_0 K_e^{\frac{1}{2}}}\label{eq:cost_system_2}
}
Here we denote supply weights $w_e$ which encapsulate the effective share a supplying vessel provides, enabling further tuning of essential and redundant vessels in the adjustment process. 
To reduce complexity we set $w_e=1$, granting each vessel the same supply effectiveness.
It should be noted that in this particular model framework, each absorbing volume only needs to be in touch with at least one supplying vessel to ensure supply, in principal. 
In order to ensure the same amount of initial vessels per service volume and vice versa, we impose periodic boundaries on the system. 
For the moment, we define each independent, shortest cycle in the graph as a service volume, given a hexagonal lattice this means each volume is to be affiliated with six edges. 
In any system with periodic boundaries we impose a single source with solute influx on a random position and a single sink with absorbing boundary on one of the topologically most distant sites, see Figure \ref{fig:setup_2_links}.
We set all vessels to respond to a volume demand $\phi_{0,v}$ such that the network's demanded filtration rate would correspond to 
\g{
\sigma_0= \frac{\sum_v \phi_{0,v}}{\sum_{v, J_v>0} J_v}
}
For the presented simulations, we initialize $\phi_{0,v}$ homogeneously across the network, as we do for $\beta^*$.
As described in the previous section, we find the system's stationary states and analyze those for their nullity $\varrho$ and actual filtration rate $\sigma$, as defined in the previous section \ref{sec:orderparameters}.
Note that non-zero nullity, for periodic boundaries, may correspond to the existence of topological generators (cycles created by walking through the periodic boundaries) \cite{Modes:2016df}. \\
Once again we focus on three significant $\bra{\sigma_{0},\beta^*}$ variations as depicted in Figure \ref{fig:volume_supply_screen} and \ref{fig:volume_supply_screen_networks}.
Surprisingly we find that the nature of topological transitions and filtration regimes mostly preserved in comparison to the case of linkwise supply:
As before, we observe the formation of dilated bulk vessels with collapsing peripheries, dangling branches as well as shunting for high demand and low absorption rates, see Figures \ref{fig:volume_supply_screen}a and \ref{fig:volume_supply_screen1}. 
The emergence of flow bottlenecks seems to be a recurring motif due to the choice of Neumann boundaries.
Note, that the resulting network redundancy becomes more sensitive to the plexus random radii initialization as well as the fact that full plexus recovery becomes unlikely. 
Further we find the topological transition once again to correlate with the breakdown of filtration.
Turning toward the regime featuring reduced demand paired with increased absorption capability one may observe the emergence of re-entrant behavior, see \ref{fig:volume_supply_screen}b and \ref{fig:volume_supply_screen2}.
Nevertheless one may find the filtration demand to be met in good agreement unless increased $\alpha_0$ is considered.
Note that the redundancy during this re-entry is mostly generated due to vessel paths enclosing multiple merged tiles.
Finally, we consider the case of low demand and high absorption rates, see \ref{fig:volume_supply_screen}c and \ref{fig:volume_supply_screen3}. 
As before we notice the emergence of nullity re-entrance and considerable mismatch of the resulting filtration in comparison to its initial demand.
All in all, though we imposed complex uptake architecture we do not find significant changes in this regime. 
So far we have not found any demand-supply model, neither linkwise nor volume-wise, to guarantee all initially defined service volumes to stay in touch with at least one supplying edge.

\section*{Discussion}
\section{Discussion}\label{sec:discussion}
In this study we have shown that demand-supply driven remodeling can be severely perturbed by wall-shear stress driven adaptation and vice versa. \\
First, given the particular demand-supply mismatches $S\bra{\vc{\Phi},\vc{\Phi}_0}$, e.g. weakly absorbing vessels paired with high demand service volumes, we have shown the existence of a nullity transition in combination with the formation of frustrated structures, such as bottlenecks and dangling branches. 
Further we observed the emergence of nullity re-entrant behavior for strongly absorbing, low demand service volumes, and find its existence for both link- or volume based mismatch scenarios $S\bra{\vc{\Phi},\vc{\Phi}_0}$. 
Neither periodic boundaries nor the re-structuring of injection points seem to alter the qualitative behavior of the adaptation process, i.e. the occurrence of topological transitions.
Generally we find the establishment of homogeneous uptake patterns in the network for negligible dissipation feedback $\alpha_1$, see supplement S\ref{apx:extra_plots}, in accordance to studies on purely metabolite optimized systems \cite{Meigel:2018hw}.
Overall we find wall-shear stress based adaptation and metabolite uptake optimization to be competing mechanisms leading to hydrodynamically unfavorable vessel formations, when absorption rates $\beta^*$ are poorly adjusted with regard to the supply-demand mismatch $S\bra{\vc{\Phi},\vc{\Phi}_0}$.
It seems unlikely though that real biological systems operate in this regime unless forced to do so due to pathological ramifications.
One may argue that these phenomena arise due to the shortcomings of the model framework, as we consider (in-compressible) Hagen-Poiseuille flow with Neumann boundaries.
Due to these model characteristics arbitrarily large pressure gradients and shear stresses may be present without leading to catastrophic failure of the actual network.
At the very least it indicates a limit of the applicability to Kirchhoff network based adaptation schemes.
As was demonstrated in \cite{Meigel:2018hw,gavrilchenko2020distribution}, optimizing a flow network for homogeneous solute uptake alone will not spark topological transitions or pruning events whatsoever.
Yet we do not find the metabolite uptake mechanism to ensure stabilization of space-filling, robust vessel systems in case of increased volume penalties.
As a matter of fact, we found that without any dissipation feedback considered, large network sections can collapse and fracture.
This seems to hold true for either demand-supply scenario, as even in the case of volume supply  we do not find the uptake mechanism to necessarily stabilize the minimal amount of vessels in contact with each service volume. 
Future studies should therefore consider diverging penalties in case of degeneration of all affiliated vessels to avoid catastrophic outcomes. This may point to a need to consider strongly non-linear dynamics in the regulatory networks that ultimately control these penalties effectively.\\
Assuming homogeneous absorption rates and demand, we find a transition between metabolic zonation scenarios, as previously discussed by Meigel et al \cite{Meigel:2018hw}, see supplement S\ref{apx:extra_plots}.
Hence, in comparison to real capillary systems, we find our model to account for homogeneously supplied structures, e.g. as found in zebrafish vasculature, where flow uniformity was considered as a crucial factor in development \cite{Chang:2019ipa,2017PLSCB..13E5892C}. 
On the other hand, it appears that wall-shear stress driven adaptation worsens the demand-supply mismatch as the bulk of absorbing vessels is pruned down to a few conducting channels corresponding to a set of linear channel solutions.
The emergence of dangling branches and bottlenecks seems more relevant as these correspond to formations found pathological vessel development, e.g. in cases of tumor driven sprouting or stenosis.\\ \vspace{0.25cm}
We nevertheless come to the conclusion that the adjustment of flow landscapes via radial adaptation alone poses limited adaptation potential in order to ensure metabolite uptake and hydrodynamic efficiency simultaneously.
That is, if the demand-supply mismatch is unfavorable, which gives rise to a set of potential new questions to be solved: How effective may tissue elements regulate $\beta^*$ as to adjust local uptake given a flux, i.e. by modifying membrane porosity, density of transporters, internal enzyme levels for clearance or storage capabilities.
How effective are tissues able to regulate their responses and readjust volume penalties and dissipation feedback for quickly altering metabolite demands, e.g. in case of organism growth or tumor angiogenesis?
Are $\sigma_0$, $\beta^*$ levels regulated by tissue elements in order to prevent frustrated formations and is the appearance of unfavorable demand-absorption rates symptom to a pathological state?\\ \vspace{0.25cm}
Hence we think that the current models of antagonistic adaptation mechanisms, flux or demand-supply based, are in need of further improvement:
In terms of cost-optimization models we propose a leap-frog-style adaptation where a solute uptake optimization follows a conventional dissipation-volume optimizing system. 
Considering the conventional stress driven adaptation schemes with cost rescaling \cite{Bohn:2007fi} and stochastic flow patterns \cite{Corson:2010ee, Ronellenfitsch:2019fea} one may still generate topological complex, space-filling structures, or reach those by different dynamic environments \cite{Ronellenfitsch:2016hh,PhysRevResearch.2.043171}.
Yet, instead of readjusting radii directly given the tissue's metabolic demands in terms of oxygen, glucose etc., we suggest that once a space-filling perfusion is reached, a secondary optimization takes place adjusting $\beta^*$.
Unlike changing the flow pattern, and therefore clashing with shear stress adaptation, regulation of $\Phi_e$ would take place locally at the membrane-tissue interface.
We think this ansatz to be promising in particular in the case of complex embedded networks with metabolic zonation, such as found in the liver lobule \cite{Gumucio}, where elaborate membrane dynamics and clearance mechanisms ensure the transport of metabolites between multiple flow networks and mediating cell layers \cite{SiTayeb:2010bl}.
It has further been argued that the metabolic dynamics in such a system would be dependent on the concentration gradients of various metabolites \cite{Berndt:2018bn}, which would correspond to sophisticated $\beta^*\bra{c}$ in our current framework.
Eventually we envision this class of models to create a better understanding of the formation and maintenance of these complex intertwined systems, ultimately illuminating the relevant transport mechanism in these organs.

\section*{Supporting information}


\paragraph*{S1 Appendix.}
\label{apx:metabolite_transport}
{\bf Metabolite transport in arbitrary Kirchhoff networks.} Derivation and summary of relevant steps for flux computation in Kirchhoff networks.

\paragraph*{S2 Appendix.}
\label{apx:demand_supply_adaptation}
{\bf Demand-supply based adaptation algorithm.} Derivation of Jacobians and other matrix components for gradient-descent evaluation.

\paragraph*{S3 Appendix.}
\label{apx:extra_plots}
{\bf Concentrations, uptake and radial patterns in optimized flow networks.} Additional material on concentration, effective uptake and radial distributions in linkwise optimized flux networks.

\section*{Acknowledgments}
Felix Kramer gratefully acknowledges support from the German Federal Ministry of Education and Research (BMBF), Grant no. 031L0044 (SYSBIO II). Our thanks go to Felix Meigel, Karen Alim, Benjamin Friedrich, Marino Zerial and Yannis Kalaidzidis for their insightful feedback during the development and implementation of the model, as well as for their constructive feedback regarding the manuscript. Further thanks go out to the members of the Modes Lab, Zechner Lab, Huch Lab, Honigmann Lab and Grapin-Botton Lab of the CSBD and MPI-CBG.\\
In the course of preparing this manuscript we became aware of related unpublished work by Gounaris et al \cite{Gounaris:2021}, concurrently posted on the Arxiv.
\nolinenumbers

%
%
%

\end{document}